# Understanding Mode Choice Behavior of People with Disabilities: A Case Study in Utah

[1]Megh Bahadur KC, [2]Dr. Ziqi Song*, [2]Dr. Keunhyun Park, [3]Dr. Keith Christensen

[1, 2*]*Department of Civil, Structural and Env Engineering, University at Buffalo, SUNY, 233 Ketter Hall, Buffalo, NY, 14260*

[2]*Department of Forest Resources Management, The University of British Columbia, Vancouver, BC*

[3]*Department of Landscape Architecture and Environmental Planning, Utah State University, Logan, UT, USA*

Email: [1]meghbaha@buffalo.edu, [2*]zqsong@buffalo.edu, [2]keun.park@ubc.ca, [3]keith.christensen@usu.edu
ORCID: [1]0000-0001-5257-5779, [2*]0000-0002-9693-3256, [2]0000-0001-5055-7833, [3]0000-0001-8455-1456
*: *Corresponding author*

**ABSTRACT**

Despite the growing recognition of the importance of inclusive transportation policies nationwide, there is still a gap, as the existing transportation models often fail to capture the unique travel behavior of people with disabilities (PwD). This research study focuses on understanding the mode choice behavior of individuals with travel-limited disabilities and comparing the group with no such disability. The study identified key factors influencing mode preferences for both groups by utilizing Utah's household travel survey, simulation algorithm and Multinomial Logit model. Explanatory variables include household and socio-demographic attributes, personal, trip characteristics, and built environment variables. The analysis revealed intriguing trends, including a shift towards carpooling among disabled individuals. People with disabilities placed less emphasis on travel time saving. A lower value of travel time for people with disabilities is potentially due to factors like part-time work, reduced transit fare, and no or shared cost for carpooling. Despite a 50% fare reduction for the disabled group, transit accessibility remains a significant barrier in their choice of Transit mode. In downtown areas, people with no disability were found to choose transit compared to driving, whereas disabled people preferred carpooling. Travelers with no driving licenses and disabled people who use transit daily showed complex travel patterns among multiple modes. The study emphasizes the need for accessible and inclusive transportation options, such as improved public transit services, shorter first and last miles in transit, and better connectivity for non-motorized modes, to cater to the unique needs of disabled travelers. The findings of this study have significant policy implications such as an inclusive mode choice modeling framework for creating a more sustainable and inclusive transportation system.

**Highlights**

- Use of revealed preference statewide household survey, map APIs, and simulation algorithm for discrete choice modeling
- MNL mode choice models specific to people with disabilities, people without disabilities, and the general population in Utah
- Despite a 50% fare reduction for PwD, transit accessibility remains a significant barrier in their choice of Transit mode
- An inclusive mode-choice modeling framework is required to address the unique travel needs of PwD
- Potential policy implications for inclusive and equitable transportation system planning in Utah and beyond

# 1. INTRODUCTION

Transportation is essential to modern society, enabling individuals to access various activities and services vital to their daily lives. For many people, the transportation mode choice can significantly impact their mobility and overall quality of life. Among these travelers, a significant portion, 8-10% of the population, includes people with disabilities (PwDs) (Bureau of Transportation Statistics [BTS]) (BTS, 2018). Moreover, nearly one in five Utahns lives with a disability (*UDoHHS*, 2017), whose statistics come to 9.2% for those with mobility-limiting disabilities in working-age adults, as per the Utah Department of Health (*IBIS*, 2021). This significant portion of the population has unique travel patterns and needs and, thus, demands particular attention in transportation planning.

In addition, PwDs face numerous challenges while traveling for different trip purposes. Traveling outside the home can be challenging for individuals with disabilities, as their disability can pose various risks and impacts. For instance, it may hinder their ability to drive or find suitable parking independently. Despite efforts to improve transportation accessibility, individuals with disabilities face challenges in accessing appropriate transportation options. The Americans with Disabilities Act [ADA] (*ADA*, 1990), enacted several decades ago, aimed to enhance inclusivity and remove barriers. However, transportation systems still struggle to meet the needs of this diverse population (Myers et al., 2022). With an aging population and an increasing proportion of individuals with travel-limiting disabilities, addressing the travel behavior of PwDs has become a growing concern for Western countries the most (Rosenbloom, 2001).

The existing transportation models often fail to capture the unique travel behavior of PwDs, leading to inaccurate forecasts, poor decision-making, and transportation inequity. Moreover, PwDs are prone to slow transportation modes while making short-distance trips (Park et al., 2023). Furthermore, disability is often associated with poverty, affecting vehicle ownership and travel behavior, further exacerbating transportation disparities (Okoro, 2018). Additionally, transportation accessibility can be particularly challenging in rural areas compared to urban regions, making travel even more limited for PwDs (BTS, 2018). While inclusive transportation is increasingly acknowledged as crucial, a notable disparity remains due to the inadequacy of current transportation models in representing the distinct travel patterns of individuals with disabilities in mode choice studies (Park et al., 2023). Hence, the mode choice models designed for the general population may not be suitable for this group and need specific studies on mode choice behavior.

To address the gaps in understanding and accommodating the travel behavior of PwDs, this research study focuses on developing discrete mode choice models specific to this population group and comparing the results with no disability group in Utah. By utilizing revealed preference (RP) household travel survey data [HHTS 2012] (Utah Travel Study, 2013) that have information on various household, individual, work-related, built environment, and travel-related characteristics, this study aims to identify critical parameters influencing the mode choice behavior of individuals with disabilities. Through simulation algorithms and data preprocessing using the Google Maps API, in addition to the HHTS 2012 provided data, this research intends to enhance the rigor and robustness of the conventional discrete mode choice models.

The MNL model has been selected for its simplicity, robustness, low technical threshold, easy implementation, and generality. MNL holds the problem with more detail and reasonable factor levels for multiple explanatory variables (Zhao et al., 2019). There is a growing trend of incorporating non-motorized modes such as walking and biking (Cook et al., 2022; Lawton, 1989; Purvis, 1997). Utah has demonstrated and planned many more active transportation projects throughout the state (Peay et al., 2020), adding more diversity in travel mode choices. Therefore, it is worth including both motorized and non-motorized modes in our analysis.

By shedding light on the travel behavior of PwDs, this study will contribute significantly to inclusive transportation planning and policy-making in Utah and beyond. We develop mode choice models separately for the general population, people without disabilities (Pw/Ds), and PwDs. Interpreting their results and making comparisons specifically tailored to PwDs will provide valuable insights into their unique needs and preferences. Moreover, identifying the factors influencing mode choice behavior for individuals with disabilities can guide the design of targeted policy interventions to promote equitable and sustainable transportation. Through this research, transportation planners and policymakers understand the complex dynamics of the mode choice behavior of PwDs.

## 2. LITERATURE REVIEW

Transportation is an essential component of daily life for many individuals, including PwDs. Commute mode shares in the US constitute 76.6% for drive alone, 9% for carpooling, 5.2% for transit, 3.4% for walking and bike mode, and 1.2% for other modes (BTS, 2018). Utah alone counts 75.7%,12.4%,2.4%,3.6%, and 1.1% for driving alone, carpooling, transiting, walking and biking, and other modes, respectively (*U.S. Central Bureau of Statistics*, 2016). Moreover, people's trip-making behavior varies significantly by age group, employment status, and mobility restrictions. BTS (2018) reported that 25.5 million (8.5% of the total population in the USA) Americans aged five or older had travel-restraining disabilities. Transportation-handicapped people in rural Virginia highly chose the paratransit system, which provides door-to-door service (Stern, 1993).

PwDs have unique travel patterns. Failure to address the travel behavior of individuals with disabilities results in inaccurate forecasts and poor decision-making and further amplifies transportation inequity for this disadvantaged population (Park et al., 2023). PwDs experienced limited access compared to those without a travel-limiting disability. Although the ADA was enacted a few decades ago, transportation is still a barrier to full inclusion (Myers et al., 2022). In addition, PwDs face numerous challenges while traveling for different trip purposes, such as work, shopping, or other trips. In general, PwDs tend to make fewer trips than those without disabilities, such as work trips and social/recreational activities (Myers et al., 2022; Park et al., 2023), which may lead to erroneous results for estimating transportation systems.

Disability is prevalent in poverty, connected to vehicle ownership and limiting trip-making behavior (Okoro, 2018). Transportation accessibility in rural places is more complicated than in urban or suburban areas. For example, using transit services by disabled people in rural areas could be limited or even impossible. Traveling outside the home can be challenging for individuals with disabilities, as their disability can have various risks and impacts. For instance, it may hinder their ability to drive or find suitable parking independently. Navigating public transportation can also pose difficulties, and they may encounter obstacles in places with environmental barriers. Although there is increasing awareness about the significance of inclusive transportation policies, a notable gap remains in addressing the distinct travel patterns of individuals with disabilities within the realm of mode choice modeling literature (Park et al., 2023).

A significant portion of recent literature concentrates on utilizing RP survey data for the purpose of mode choice modeling. Ding et al. (2015) studied the choice model using RP surveys in Maryland. They saw the effect of a change in departure time for transit and concluded that transit commuters are more sensitive to changes in travel-related attributes than car commuters. Bastarianto et al. (2019) conducted a tour-based mode choice model in Indonesia using secondary RP data and traditional discrete choice modeling (DCM). Ding and Zhang (2016) used joint RP and stated preference(SP) surveys for traveler categorization using cluster-based analysis by taking into account the traveler's characteristics. However, data collection was only from a specific place, hindering the results' generalizability. All these studies partially used the RP survey, and the DCM implementation was only for the overall population dataset without segregation of specific groups. The trend of using RP data might be attributed to the benefits RP data offer in accurately capturing the real travel behaviors exhibited by participants (Tabasi et al., 2023).

There have been some studies on specific groups of people and particular locations and comparing their results, including disability. Myers et al. (2022) used national household survey data [NHTS 2017] to study the travel patterns of adults with travel-limiting disabilities. They grouped the dataset for rural and urban

America using a simple logistic regression model and revealed that rural non-drivers were less likely to take trips. They did not capture a wide range of travel, socio-demographic, and built-environmental variables. Du et al. (2021) used an MNL model to examine travel mode choices among healthcare-seeking travelers, comparing elderly and non-elderly patients based on various healthcare-related surveys. They found that with increasing age, car travel decreases while non-motorized modes increase for non-elderly individuals but exhibit reverse behavior in elderly individuals. This study does not capture disability attributes in the overall trip scenario.

In a similar study conducted in China on mode choice for urban elderly healthcare activity, a comparison between geographically distinct areas (core area and suburban area) revealed that hospital service improvements could incentivize elderly individuals to opt for green modes in the core area. In contrast, those seeking healthcare in the suburban area prefer bus travel due to longer distances (M. Du et al., 2020). They also limited their study to age, gender, and income variables, whereas including vehicle ownership, employment, bike availability, and built environment variables could have improved their model. Henly and Brucker (2019) used the NHTS 2017 dataset to identify travel patterns of working-age Americans with disabilities. They only focused on shopping trips and the length of the period during which people have a disability.

The mode choice behavior study can treat land use and transportation system characteristics as exogenous variables (Pinjari et al., 2007). Controversies exist regarding the effect of built environment characteristics on the mode choice model. Travel variables are generally inelastic concerning changes in built environment variables and population densities that weakly associate with travel behavior (Ewing & Cervero, 2010). However, road network density, walkability index, and residential location proximity to transit stations positively affect the non-motorized mode choices (Clark & Scott, 2014).

Moreover, traditional models have mainly encompassed motorized modes of transport, and there is a growing trend of incorporating non-motorized modes such as walking and biking (Cook et al., 2022; Ewing et al., 2019; Lawton, 1989; Purvis, 1997). Utah has demonstrated and planned many more active transportation plans and projects throughout the state (Peay et al., 2020), adding more diversity in travel mode choices. Ewing et al. (2019) reported over 50% of the Utah metropolitan planning organizations (MPOs) have not included non-motorized modes in their mode choice modeling.

There has been no research utilizing RP survey data using map APIs to analyze the mode choice tendencies of individuals with and without disabilities by including non-motorized modes. Based on the literature review above, none of the studies fitted the choice model for household, person-related, trip-related, and built environment characteristics for PwDs. In addition, we are able to use RP survey data for mode choice modeling and capturing the deficiencies in the travel behavior literature for disabled people. Our research strengths are capturing such uniqueness in travel for specific groups by incorporating non-motorized modes in mode choice modeling and comparing the diverse models. The mode choice models developed in this research will provide a robust framework for analyzing individual trip-level data, generating insights and enhancing existing practical mode choice models that can inform transportation planning, policy-making, and sustainability efforts.

## 3. METHODOLOGY

### 3.1. Data

We used RP survey data from Utah's 2012 household travel survey. It includes a statewide household travel diary to support regional and statewide transportation planning. The HHTS includes whether a respondent has a travel-limiting disability or not. Those who reported affirmative are known as having a travel-limiting disability or just a "disability." This is the uniqueness of our dataset to analyze the impact on travel behavior for PwDs. Household, person, and trip-level data were processed and weighted to represent the true population in the state. The final dataset includes 18,171 adults, 8,875 children, and 101,404 trips.

The data were arranged separately for households, persons, vehicles, and built environments. We merged trip trip-level data with the person-level data. Built environments and vehicle data sheets are combined with households. Then, the final dataset is prepared, which includes household and demographic characteristics, personal and travel attributes, built environment factors, and vehicle characteristics. Household weights are also provided to make the sample representative of the Utah population. More details about weighting can be found in the Utah Travel Study (2013).

### 3.2. Data Preparation and Analysis

The big picture of this research is MNL mode choice modeling of four alternative modes (i.e., drive alone, carpooling, transit and nonmotorized). The choice set has been defined, including sociodemographic attributes, travel attributes, mode attributes and built environment attributes. Different datasets on household characteristics, person characteristics, the built environment, and vehicle datasets were gathered and imported into the Python pandas library. The County traffic analysis zones (coTAZID) files were merged to make a master dataset. The unreported trip distances and long trip distances are removed from the dataset, assuming they were outliers. Using this RP survey for understanding mode choice modeling would require each mode-specific travel time for each trip, although there is self-reported travel time for the primary mode. The RP survey reported only the chosen mode and travel time for that mode. We need to know what the travel time would have been the travel time if other available modes had been chosen during the surveyed period. Techniques like the Google Maps directions API (Tabasi et al., 2023) have the capability to retrieve data about these options. However, this procedure might involve intricacies and consume a significant amount of time. For non-motorized modes, the traveling speed for walking is assumed to be 2.5-3 miles per hour, whereas the biking speed is 10 miles per hour (Alves et al., 2020) and the weighted average is executed.

Furthermore, adding the trip cost in accordance with the mode would have a significant effect on mode choice modeling, thereby making the modeling rigorous and robust. The author developed Bing API, Google Maps API, and a simulation algorithm to extract attributes like travel cost that weren't easily perceptible or self-reported. Similar methods can be found in Tabasi et al. (2023). Travel cost has been calculated by assuming the car fuel cost, average insurance, average maintenance and operation cost, and transit fares. Transit use is discounted at 50% for those who have a disability. Liao et al. (2020) reported that public transit has about twice the travel duration of car travel. Different data visualization sets, categorical variable mapping, and one-hot encoding for categorical variables are performed. We performed Chi-Square tests to compare the variables (Myers et al., 2022) between the two groups (people who do not have a travel-limiting disability and people who have such a disability), and statistical differences were also reported. Four major travel modes are drive-alone, carpool, transit, and non-motorized. Auto travel with two or more occupants without a driving license is designated as carpool trips.

Mode-specific availability is defined; for example, drive-alone is available to those whose household has at least one car and a driving license. Carpool is available for all. Transit is available except that the person explicitly identifies never to be used. Non-motorized mode is available if the household has at least one bike or walking distance is less than 5 miles. With these selections, 68,842 trips were deemed feasible for the general population, Pw/Ds for 67,505 trips, and 1,337 trips for PwDs. Zhou et al. (2022) have used 114,847 trip data to explore the mode choices of older and younger adults.

### 3.3. Modeling

The random utility maximization approach (McFadden, 1974) was adopted to analyze the mode choice behaviors of people with and without disabilities, where alternatives with the highest utility are chosen among the provided alternatives (Marschak, 1959). Utility refers to an indication of value to an individual. The utility value U (with its defined set of functions) decides which alternative to choose from all available sets of alternatives. The utility for any decision maker involves the deterministic or observable component, another component that cannot be directly measured, or an unknown utility component. This unobservable component is called random error in utility (Koppelman & Bhat, 2006; McFadden, 1974).

We explore and compare the determinants of mode choice for trip-making by the general, non-disabled, and disabled populations. The MNL model is proposed for data analysis and travel behavior study for its simplicity, robustness, low technical threshold, easy implementation, and generality (M. Du et al., 2020). MNL processes multiple equations for the analysis of k-categories of the dependent variable to multiple independent variables (continuous and categorical at the same time) through logistic equations (M. Du & Cheng, 2018; McBain & Caulfield, 2018). Panda Biogeme v3.2.11 (Bierlaire, 2020) on Windows 10 is used for data analysis. In MNL models, Drive alone (car), Carpool, Transit, and Non-motorized (walk and bike) are dependent variables. The accuracy of the MNL model is established with McFadden and Chi-square (−2*Log-likelihood values for the equal probability model minus the corresponding value for the final MNL model) values. Estimating the value of travel time (VOTT) provides insights into the sensitivities of mode choice concerning travel time and travel cost. Tabasi et al. (2023) highlighted that estimating willingness to pay (WTP) through diverse methods, data sources, and studies can aid in selecting a dependable WTP value (or range) for policy evaluation, as even a slight alteration in WTP can significantly influence analysis outcomes. To calculate VOTT or WPT, we obtain the ratio of parameter estimates for travel time and travel cost from the utility equation $VOTT = \frac{\beta_{Time}}{\beta_{Cost}}$ (Koppelman & Bhat, 2006). VOTT parameters are kept fixed for all individuals in our three models.

### 3.4. Model Specification

The car (driving alone) is used as a reference mode, and alternative specific constants (ASCs) were defined for each mode. The travel time and travel costs are generic constants affecting each mode's utilities. Disability coefficients are defined to represent the presence of the disability effect in our model. The gender-specific variable is female, where males are the reference. The transit frequency is added, making the mode specific to the transit mode. The built environment variable includes a residential place type. The residential places here are CBD, urban, suburban, and rural. Three categories of employment status: full-time, part-time, and no work. All these variables and their respective beta coefficients are incorporated in the MNL model specification. Some mode-specific variable terms were removed after several model runs and testing their significance.

The MNL model for the general population is regarded as model1, whereas the model for non-disabled people is represented as model2. The people who reported having travel-limiting disabilities only are taken as a model for PwDs, known as model3 onwards. The model specifications for model3 do not have the beta coefficients for a disability that model1 includes. Other model specifications remain similar so that model interpretation and comparisons can be performed easily.

# 4. RESULTS

## 4.1. Key Sample Statistics Results

*Table 1: Descriptive statistics of variables used in final MNL model*

| Models | | | Model1 | Model2 | Model3 | p-val |
|---|---|---|---|---|---|---|
| **Trip Purpose** | | ALL | 68842 | 67505 | 1337 | |
| **Total sample** | | | 100 | 98.05 | 1.95 | |
| A | **Household & Socio-demographic characteristics (%)** | | | | | |
| 1 | HH size | 1 | 6.74 | 6.50 | 19.59 | < 0.001 |
| | | 2 | 26.12 | 25.86 | 40.57 | |
| | | 3 | 14.14 | 14.10 | 16.29 | |
| | | 4 | 17.36 | 17.52 | 8.30 | |
| | | 5 | 16.83 | 17.00 | 7.38 | |
| | | 6 | 18.82 | 19.02 | 7.87 | |
| | HH income | Low | 14.47 | 14.20 | 29.16 | < 0.001 |
| | | Med | 13.56 | 13.61 | 10.68 | |
| | | High | 41.32 | 41.64 | 23.55 | |
| | | Very high | 20.32 | 20.48 | 11.41 | |
| | | NA | 10.34 | 10.08 | 25.20 | |
| 3 | Veh Owner ship | No auto | 0.55 | 0.44 | 6.77 | 0 |
| | | 1 | 15.60 | 15.27 | 33.68 | |
| | | 2 | 53.13 | 53.39 | 38.62 | |
| | | 3+ | 30.73 | 30.90 | 20.93 | |
| 4 | # HH workers | 0 | 13.34 | 12.85 | 40.51 | < 0.001 |
| | | 1 | 46.69 | 47.00 | 29.23 | |
| | | 2 | 33.55 | 33.80 | 19.65 | |
| | | 3+ | 6.42 | 6.35 | 10.62 | |
| 5 | # Bikes for Adults in HH | 0 | 26.16 | 25.81 | 46.19 | < 0.001 |
| | | Upto 3 | 52.44 | 52.68 | 38.99 | |
| | | 3 plus | 21.40 | 21.51 | 14.83 | |
| D | **Built Environment Characteristics (%)** | | | | | |
| 1 | Residential place type | CBD | 2.83 | 2.75 | 7.26 | 0.392 |
| | | Urban | 24.77 | 24.73 | 27.15 | |
| | | Suburban | 17.38 | 17.41 | 15.38 | |
| | | Transition | 34.05 | 34.22 | 25.02 | |
| | | Rural | 20.97 | 20.89 | 25.20 | |

| | | | Model1 | Model2 | Model3 | p-val |
|---|---|---|---|---|---|---|
| B | **Traveler characteristics (%)** | | | | | |
| 1 | Gender | Male | 46.80 | 46.85 | 43.50 | 0.295 |
| | | Female | 53.20 | 53.15 | 56.50 | |
| 2 | Age | Children | 23.59 | 24.01 | 0.00 | < 0.001 |
| | | Adult | 65.04 | 64.92 | 72.18 | |
| | | Old (65+) | 11.36 | 11.07 | 27.82 | |
| 3 | Limited mobility | Yes (1) | 1.76 | - | - | - |
| | | No (2) | 74.65 | | | |
| | | NA (0) | 23.59 | | | |
| 4 | Driving license | Yes | 75.43 | 75.43 | 75.17 | < 0.001 |
| | | No | 1.73 | 1.32 | 24.83 | |
| | | NA | 22.84 | 23.25 | 0.00 | |
| 5 | Emplo yment | Full-time | 32.83 | 33.07 | 19.46 | 0.023 |
| | | Part-time | 12.57 | 12.51 | 16.11 | |
| | | No | 28.63 | 28.32 | 46.00 | |
| | | NA | 25.97 | 26.10 | 18.43 | |
| C | **Trip characteristics (%)** | | | | | |
| 2 | Main modes | Drivealone | 47.05 | 46.98 | 41.90 | < 0.001 |
| | | Carpool | 36.01 | 36.18 | 48.52 | |
| | | Transit | 8.48 | 8.36 | 1.58 | |
| | | NM | 8.46 | 8.48 | 8.00 | |
| 3 | Transit frequency | Daily(cat1) | 1.57 | 1.46 | 7.69 | < 0.001 |
| | | Few times a week (2) | 2.00 | 1.92 | 6.59 | |
| | | Few times a month(3) | 20.16 | 20.18 | 18.97 | |
| | | Never (4) | 52.68 | 52.42 | 66.75 | |
| | | NA (0) | 23.59 | 24.01 | 0.00 | |

p-values (significance) reported for Chi-Square tests between Disability and Without Disability columns.

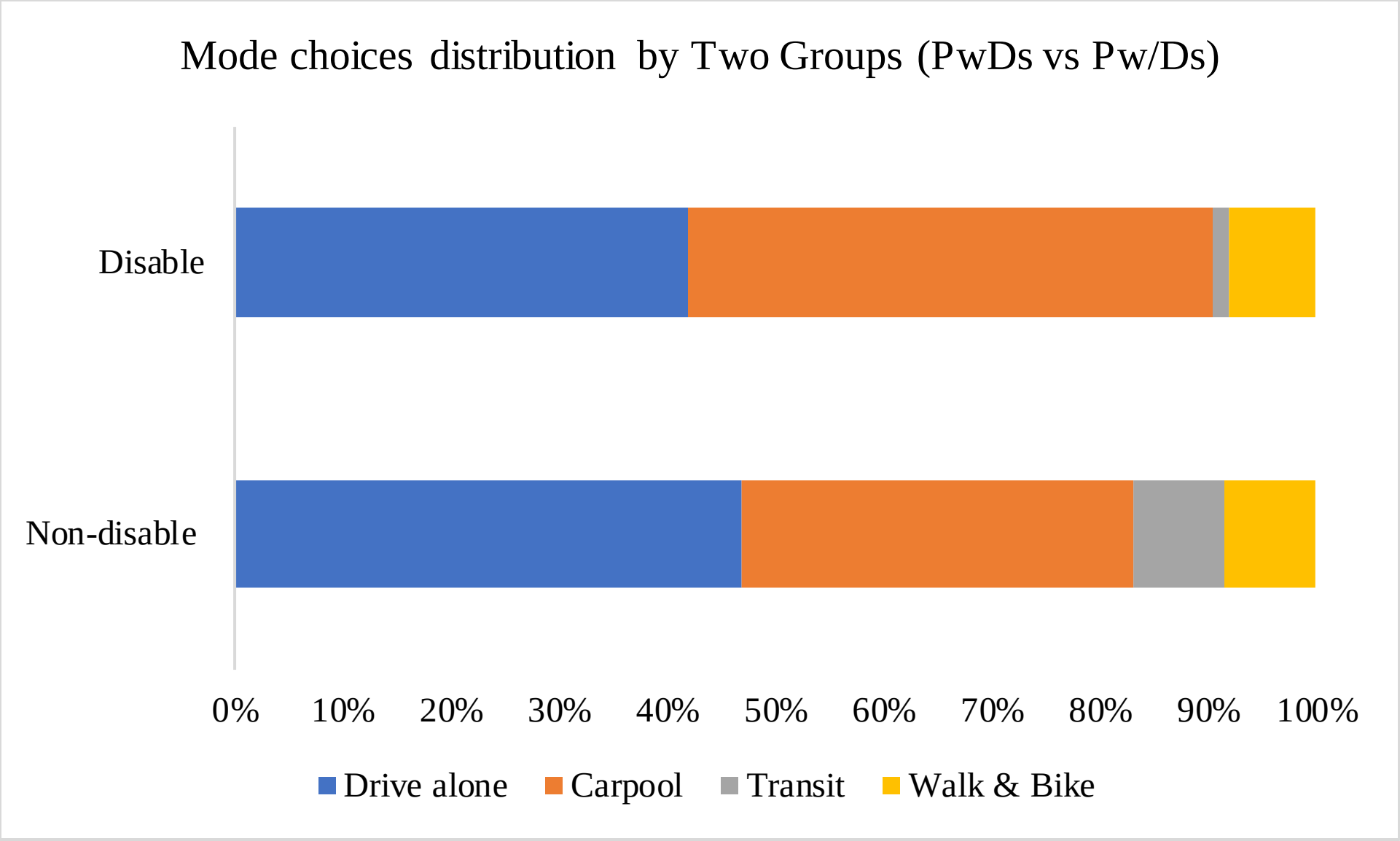

*Figure 1: Mode share distribution among People with and without disabilities*

Descriptive statistics show that almost half (47.9%) of the population who do not have disabilities use the drive-alone mode. In contrast, this share drops to 41.9% for people with limited mobility. Furthermore, PwDs use carpooling 25% more than the counterpart group in Utah, which aligns with the findings of the report prepared by BTS (2018). Transit users in the disability group seem very low compared to the non-disabled group, which is shown in Figure 1. It can be because of the inaccessible public transportation in Utah. People who have travel-limiting disabilities do not work full-time very often, and vice versa. Part-time working employees for disabled people account for 9% more than non-disabled people. Descriptive statistics on the considered variables in our model are depicted in Table 1.

PwDs take a shorter trip on average than people no such disabilities, which is 6.76 miles versus 5.6 miles for driving alone. Non-disabled people take transit for longer trips of about 11.52 miles, whereas disabled people take transit for 8.46 miles. Carpooling is used most by PwDs and for trips distance. Disable people who walk shorter than those in the counterpart group.

People with travel-limiting mobility tend to make fewer trips than Pw/Ds, and the same for the trip-making probability (Park et al., 2023). In the dataset we used, 13.5% fewer trips are taken by PwDs for home-based work trips, being 4.09 versus 4.73 trips per day. This figure rises to 40% for home-based school trips, whereas the non-home-based non-work trips frequency gap was reduced to only 6.2%. On average, people with disability in Utah had 18.6% fewer trips overall. However, the review paper of Park et al. (2023) reported that this figure would be around 25%. This shows that the travel behavior of Utahns is different from the national scenario and hence adds to the value of this study. For individual modes, drive-alone shares almost equal trip frequency for disabled and non-disabled travelers, being 5.76 vs. 6 trips per day. The trip frequency disparity is highest among transit modes, where disabled people make 4.21 trips compared to 5.71 non-disabled people. The trip distance and trip frequency comparison of the two groups is shown in Figure 2.

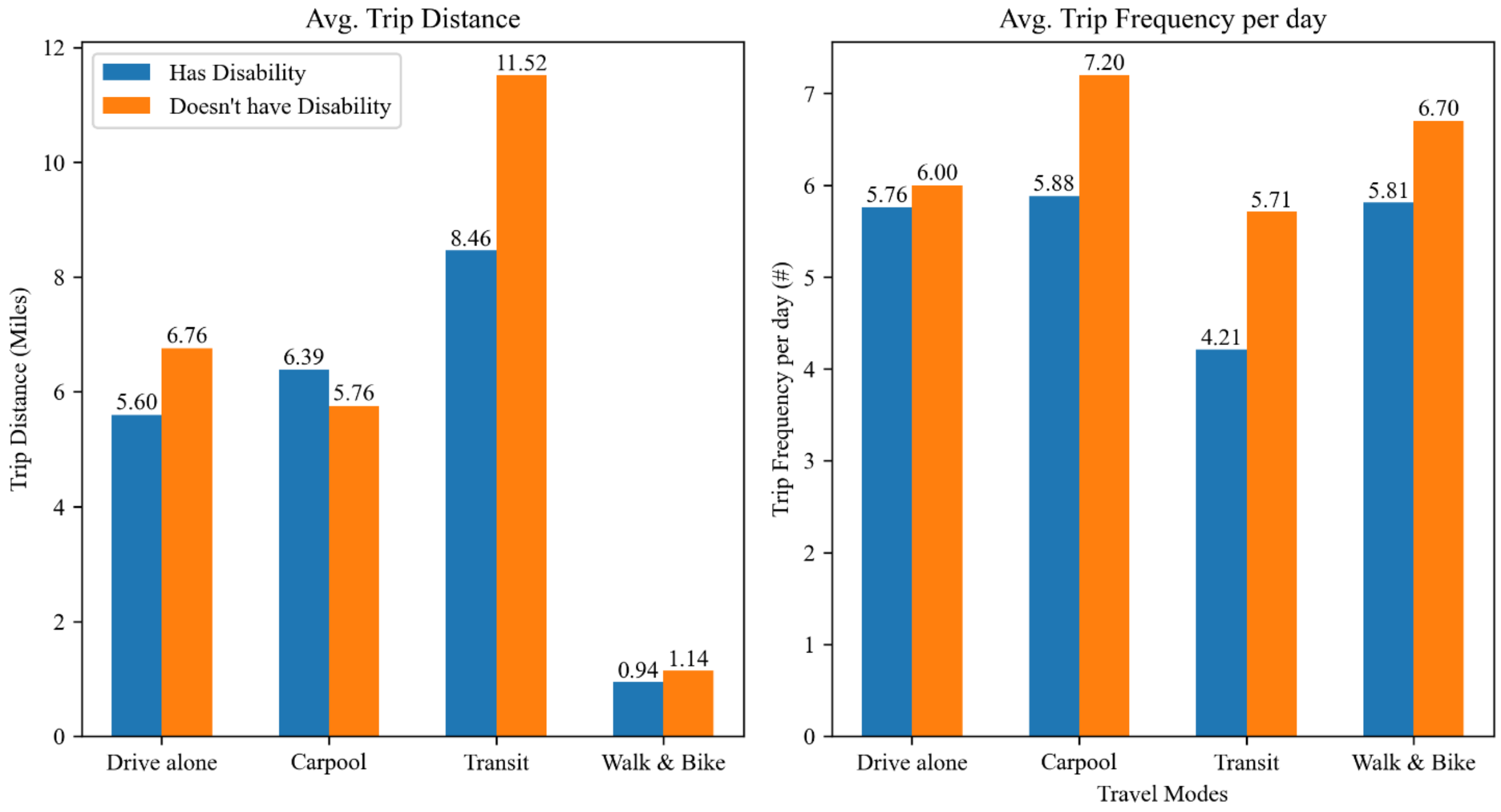

*Figure 2: Average trip distance and trip frequency comparison between models for with and without disabilities*

## 4.2. MNL Modeling Estimation Results

We performed the MNL model for our three different groups of model analysis. The final log-likelihood seemed to be improved in comparison to the null log-likelihood. Several multinomial logit models with various model specifications and parameters were fitted to attain the study objectives. To compare the estimated model, $-2LL = -2\ (LL_{base} - LL_{estimated})$ from the model results are calculated, and $\chi^2$ value is computed from $\chi^2$ distribution table. In our case, the 83 parameters are estimated. Three base constants in the model make the degree of freedom 83-3 = 80. With a 1% level of significance, the $\chi^2$ value evaluated is 112.329, which is less than $-2LL = -2(-1667 + 1061) = 1212$ . Therefore, the model for PwDs improved compared to the base model. The likelihood ratio test value of 48,122 for Pw/Ds approves the model specification and correctness. The adjusted McFadden rho-squared ($\rho^2$) value of 0.31, 0.305, and 0.315 for the three respective models signifies good prediction. McFadden recommended $\rho^2$ values of 0.2 - 0.4 should be taken to express a very good fit for the model (Louviere et al., 2000). Selecting inclusive variables involves balancing considerations of practical significance and statistical efficiency. This section depicts various models, results, interpretations, and discussions hereafter.

*Table 2: MNL model results for VOTT calculation for various models for people with and without disabilities*

| **Variable** | **Model1- General Population** | | | **Model2- People without Disabilities** | | | **Model3- People with disabilities** | | |
|---|---|---|---|---|---|---|---|---|---|
| | **Carpool** | **Bike & Walk** | **Transit** | **Carpool** | **Bike & Walk** | **Transit** | **Carpool** | **Bike & Walk** | **Transit** |
| | Ref mode: Drive alone | | | Ref mode: Drive alone | | | Ref mode: Drive alone | | |
| ASC | 1.21 *** | 0.959 *** | - | 1.24 *** | 0.959 *** | - | - | -0.031* | - |
| Time | -0.0458 *** | | | -0.0458 *** | | | -0.0515 *** | | |
| Cost | -0.00233 *** | | | -0.00225 *** | | | -0.0088 *** | | |
| VOTT ($/hr) | 11.80 | | | 12.21 | | | 3.50 | | |

#### 4.2.1. ASC and VOTT

ASCs (alternative specific constants) for the carpool, transit, and non-motorized (walk and bike) modes are calculated, given that the drive-alone mode is a reference. All else being equal, individuals in Utah state tend to have a higher preference for carpooling ($ASC_{Carpool} = 1.21$) and non-motorized mode (Hess et al., 2010; Tabasi et al., 2023). However, PwDs show no significant comparison for carpooling, but a negative preference for non-motorized mode, as the estimated coefficients suggest. The generic coefficients for time and cost $\beta_{Time} = -0.0458\ and\ \beta_{Cost} = -0.0023$ indicates that an increase in one-minute travel time is associated with a decrease in utility across all transportation modes by 0.0458. The general interpretation is that travelers prefer shorter travel time, which applies to all three models. Similarly, the cost has a negative effect on mode choice utility and tends to choose modes with lower costs in all models.

The value of travel time (VoTT) for the general population in Utah of 19.9 cents per minute suggests that each hour of travel time saved accounts for $11.80. However, the average hourly wage rate for the hourly rate job category in Utah as of 2023 is $17.28 (*Hourly Rate Salary in Utah*, 2023). That means the consumer price inflation index from 2012 accounts for VOTT, which is almost similar to the wage rate. Therefore, travelers in Utah may have a strong consideration for the time saved during travel, more than that they spend working and earning wages (Koppelman & Bhat, 2006; Tabasi et al., 2023). In contrast, the disabled population model shows lower VOTT ($3.50 per hour) among all models. One possible interpretation would be that the non-disabled population has more full-time work and must value travel time more. On the other hand, disabled people mainly work part-time and hence have less value for travel time. Another interpretation is that in Utah, there is a 50% discounted transit fare for disabled people. And at the same time, disabled people do not drive alone often, preferring to use carpool/ shared mode, which is generally less costly. Lucas et al. (2007) reported that in Honolulu, Hawaii, elderly people place more value on transport cost than travel time, which is similar to our findings here.

*Table 3: MNL results for different models (Models 1, 2 and 3) for overall population, people with and without disabilities*

| Variable | Model1- General Population | | | Model2- People without Disabilities | | | Model3- People with disabilities | | |
|---|---|---|---|---|---|---|---|---|---|
| | **Carpool** | **Bike & Walk** | **Transit** | **Carpool** | **Bike & Walk** | **Transit** | **Carpool** | **Bike & Walk** | **Transit** |
| | Ref mode: Drive alone | | | Ref mode: Drive alone | | | Ref mode: Drive alone | | |
| **ASC** | 1.21 *** | 0.959 *** | - | 1.24 *** | 0.959 *** | - | - | -0.031* | - |
| **Disability attribute** (ref: no disability) | 0.066 * | -0.318 *** | 0.364 ** | - | - | - | - | - | - |
| **Gender Attribute** (Ref: Male) | 0.471 *** | 0.0682 ** | - | 0.467 *** | 0.067 ** | - | 0.776 *** | - | 0.799 ** |
| **Employment** | | | | | | | | | |
| Fulltime | -0.777 *** | -0.396 *** | 0.401 *** | -0.793 *** | -0.41 *** | 0.428 *** | - | 0.922 ** | - |
| Parttime | -0.408 *** | - | 0.602 *** | -0.415 *** | - | 0.624 *** | 0.139 * | - | - |
| **Age Attribute** | | | | | | | | | |
| Age 16-64 | 0.107 *** | 0.559 *** | - | 0.125 *** | 0.557 *** | - | -0.517 *** | - | - |
| **Household Income** | | | | | | | | | |
| Low (<35k) | 0.066 ** | 0.128 *** | 0.171 ** | 0.0486 * | 0.112 ** | 0.171 * | 0.901 *** | 0.602 ** | - |
| Medium (35-50 k) | 0.0494 * | 0.211 *** | - | - | 0.204 *** | - | 0.457 ** | - | - |
| Very High (>100k) | -0.181 *** | -0.075 * | -0.314 *** | -0.189 *** | -0.076 * | -0.297 *** | - | - | -0.721 * |
| **Residential Location** | | | | | | | | | |
| CBD | - | - | 0.732 *** | - | - | 0.767 *** | 0.816 ** | - | - |
| Urban | 0.056 * | - | - | 0.051 ** | 0 | - | 0.373 ** | 0.573 * | - |
| **Licensed Driver** (Ref: Yes, I have) | 0.632 *** | 0.997 *** | 1.62 *** | 0.444 *** | 0.93 *** | 1.54 *** | 1.44 *** | 1.77*** | 2.33 *** |
| **Vehicle ownership** | | | | | | | | | |
| 0 | 1.75 *** | 2.82 *** | 1.97 *** | 1.42 *** | 2.71 *** | 1.58 *** | 1.03 *** | 0.906 *** | 1.206 *** |
| 1 | 0.314 *** | 0.549 *** | - | 0.333 *** | 0.574 *** | - | - | - | - |
| 3+ | -0.263 *** | -0.387 *** | -0.162 * | -0.26 *** | -0.38 *** | -0.197 ** | -0.41 * | - | - |
| **Household size** | | | | | | | | | |
| 1 | -1.84 *** | -0.725 *** | -0.523 *** | -1.87 *** | -0.77 *** | -0.414 *** | -1.82 *** | - | -0.33 *** |
| 2 | -0.849 *** | -0.265 *** | -0.219 ** | -0.847 *** | -0.25 *** | -0.194 * | -1.29 *** | -1.22 *** | -1.3 ** |
| 4 | 0.299 *** | - | 0.234 ** | 0.296 *** | - | 0.205 * | - | 0.948 ** | 1.44 *** |
| 5 | 0.497 *** | - | - | 0.486 *** | - | - | 1.45 *** | - | - |
| 6 | 0.549 *** | - | 0.4 *** | 0.536 *** | - | 0.378 *** | 1.75 *** | 1.04 ** | 1.48 ** |
| **Household Adult worker** | | | | | | | | | |
| 0 | 0.204 *** | 0.447 *** | 0.641 *** | 0.213 *** | 0.463 *** | 0.615 *** | - | - | - |
| 1 | 0.0548 *** | 0.206 *** | 0.364 *** | 0.049 ** | 0.206 *** | 0.374 *** | 0.505 ** | - | - |
| 3+ | -0.314 *** | - | - | -0.318 *** | - | - | -0.665 *** | -0.956 ** | -1.62 ** |
| **Transit Frequency** | | | | | | | | | |
| Daily | - | 0.542 *** | 1.19 *** | - | 0.477 *** | 1.15 *** | 2.22 *** | 2.78 *** | 3.15 *** |
| few times a month | -0.289 *** | -1.01 *** | -3.18 *** | -0.312 *** | -1.00 *** | -3.24 *** | 0.618 * | - | -2.04 *** |
| never | -0.443 *** | -1.48 *** | -3.56 *** | -0.463 *** | -1.47 *** | -3.63 *** | - | -1.56 *** | -2.69 *** |
| **Bike availability in HH for adults** | | | | | | | | | |
| 0 | - | -0.517 *** | - | - | -0.513*** | - | - | -0.635 ** | - |
| 3 or more | -0.085 *** | 0.498 *** | - | -0.09 *** | 0.505 *** | - | 0.55 ** | - | 0.925 * |
| **Level of significance (LoS) : 1% - *** ; 5% - ** ; 10% - *** | | | | | | | | | |
| **Goodness-of-fit statistics** | | | | | | | | | |
| # of parameters: | 86 | | | 83 | | | 83 | | |
| Sample size: | 68842 | | | 67505 | | | 1337 | | |
| Init log-likelihood: | -82018 | | | -80531 | | | -1667 | | |
| Final log-likelihood: | -57517 | | | -56290 | | | -1061 | | |
| Likelihood ratio test: | 49002 | | | 48122 | | | 1212 | | |
| Rho-square: | 0.31 | | | 0.31 | | | 0.364 | | |
| Rho-square-bar: | 0.31 | | | 0.305 | | | 0.314 | | |
| AIC | 115207 | | | 112746 | | | 2288 | | |
| BIC | 115993 | | | 113503 | | | 2719 | | |
| $\chi^2$ (df) | 114.09 (df=83) | | | 112.329 (df=80) | | | 112.329 (df=80) | | |

### 4.2.2. Traveler/Personal Attribute Parameters

Travelers in general are more likely to choose transit and carpool modes over non-motorized modes, as indicated by the positive and significant coefficients for transit (0.364**) and carpool (0.066*), respectively. Conversely, travelers are less likely to opt for non-motorized modes, as shown by the negative and highly significant coefficient (-0.318***). The population with disabilities tends to use carpooling more than driving alone in comparison to Pw/Ds. Transit is the most preferred mode for PwDs as the model3 transit coefficient is higher than the people without disability group. The selection of green modes (i.e., walking and biking) significantly decreases for those with mobility restrictions compared to driving alone. This seems persuasive enough, as people having mobility difficulties may not want to walk to make trips.

For individual attributes, females prefer to use carpool and non-motorized modes rather than driving alone compared to the male population. However, there was no significant evidence to draw a conclusion about transit use and car use with respect to gender. People with disability estimated that disabled females prefer to use a carpool rather than a car. It is contrasting that walking and biking became insignificant in the case of disabled people compared to people without disability. However, it supports our initial thought that disabled people prefer to use transit and carpool rather than walking.

As there was not enough child population with disabilities in our dataset, we skipped that population age group in our analysis. We made only two age groups (adults 18-64 years and the older population 65 plus). Making the older population a reference, our study showed that people who do not have mobility restrictions prefer to use non-motorized modes in addition to car use, where preference towards carpooling is also seen for this age group. With the increase in age, the proportion of Walk & Bike decreases, similar to Du et al.'s finding (2020). Using transit by age 18-64, non-disabled people could not be interpreted because of the p-value. Talking about the PwDs and age groups, using carpool was found to be less than drive-alone and older people. As disabled people grow older, they are likely to be accompanied by their family members using other modes, which may be accounted for here. On the other hand, we could not find significant differences between car use, transit, and non-motorized modes concerning both age groups, as they were insignificant at 10% CI.

Another variable we used here is driver versus non-drivers. We analyzed whether people have a driving license or not. Drivers being referenced, non-drivers having no travel limiting difficulty, are found to prefer transit mode the most, non-motorized being second and carpool as third, and all the coefficients being positive and statistically significant. Furthermore, PwDs surprisingly adopt similar results as Pw/Ds do. People with mobility restrictions and no driving license prefer transit, non-motorized, and carpooling. In all model results, mode choice behaviors of non-drivers are similar, with a significant impact for both groups preferring to use other available modes as compared to driving, which shows people with no driving licenses show multimodal travel patterns.

#### 4.2.3. Household and Socio-demographic Attribute Parameters

The impact of household income would be significant in mode choices, as income is directly connected to vehicle ownership (Palma & Rochat, 2000). The higher the income, the more inclined towards the drive-alone mode or decreased interest in non-motorized modes (Tabasi et al., 2023). The people without disability model shows that households with an income of more than 100k prefer to drive alone; for them, transit use is the least preferred. Households with less than 35k per year are more inclined towards transit and non-motorized modes, as the coefficient is positive and significant at 1% CI. It is similar to the results of Kim and Ulfarsson (2004). In contrast, very high-income people (100k+) with a disability do not want to ride transit, whereas the use of shared ride and non-motorized mode came insignificant as p-value suggests and seems to be less sensitive to household income and mode preference as they come to statistically insignificant in some cases. Disabled people having household incomes less than 35k per year prefer carpooling and non-motorized than driving a car.

Households that do not own a car strongly prefer to use other modes. In the case of model2, the most being non-motorized, then transit and carpool last. A similar observation can be seen in income. Therefore, it is evident that vehicle ownership and household income are strongly associated. Households with three or more cars would like to drive alone more than any other mode. Pw/Ds drive cars alone more often when having three or more cars in a household than two-car ownership, as all the coefficients for all modes came negative and significant. In the case of the model3, people having a disability and owning three or more cars in a household prefer driving to using a carpool at a 10% level of significance, which was quite surprising, and other alternatives became insignificant. In contrast, no vehicle ownership for disabled families leads to the use of transit and carpooling.

As we can assume that lower-income households may have used bikes for mobility, we used the number of bikes available for adults as a variable in our modeling. It has been found that non-disabled people (model2) prefer biking compared to the car while having three or more bikes, but for people with disability, it is insignificant. People with a disability still prefer carpooling or transit, irrespective of bike availability. It can be concluded that for the general population or Pw/Ds, having more bikes in a household makes non-motorized modes prevalent. However, people who have a disability would be less or not sensitive towards bike availability.

From our modeling, the effect of household size on travel mode choice for people without a disability and those with a disability has been worked out. It seems that single-person household prefers to travel by car or drive alone rather than any other available modes for the general population model as compared to households having three members. The log-odds of choosing non-motorized and transit mode over drive-alone decreases by 0.725 and 0.525 units with an increase in 1 household size as compared to household size 3 (reference category). All coefficients being negative and significant for the people without disability model suggests that driving alone is prevalent for single-person households. The same results for two-person households can be found in both models: PwDs and those without disabilities. It is contrasting that two households containing disabled people also prefer driving alone. It would be better to see if there are any interaction variables in determining why disabled households prefer the drive-alone mode. As the household size increases, the mode preference shifts from driving alone to other alternatives in all models. The household size increase means the number of autos per household member decreases.

Therefore, some household members may be forced to take other modes like carpooling, transit, or non-motorized. For PwDs and household size 6, carpooling is the most desired mode. For people without disability model, household seems to use carpool or transit, but there is no evidence for the non-motorized mode. It could be because of the longer walking distance or the unavailability of bikes. It is also seen that four people in a household having a disability prefer transit and non-motorized, and no evidence for carpool. In short, a single-person household likes to drive alone, and larger households are more inclined to choose carpool and transit over driving alone, meaning larger households may have to shift their modes as per convenience and availability. In contrast, non-motorized modes are generally less preferred across household sizes.

#### 4.2.4. Built environment Parameters

As discussed in the literature review section, few studies have controlled for residential preferences. In this study, we have used the residential location to see the effect on mode choice (motorized and non-motorized). For the residential location variable, one can think that the downtown area is costly for parking and may have congestion. In our model1 (general population), the coefficient for transit use in the central business district (i.e., downtown area) is positive and statistically significant. Reference mode being driven alone, this finding validated the general perception. The use of carpools dominated the drive-alone use in urban areas compared to the suburban regions in Utah.

Furthermore, PwDs use carpooled rides in CBDs, while carpooling and non-motorized modes are found to be significant as compared to suburban areas and car use. Transit use for PwDs was statistically insignificant, although the coefficient was positive compared to the drive-alone mode. Hence, populations with disabilities have different travel mode choices with respect to the built environment factors. Including more interacting built environment variables to support our findings strongly would be even more noteworthy.

#### 4.2.5. Trip Attribute Parameters

The MNL estimated coefficients in our model show the distinct mode choice preference across various transit use frequencies. The reference mode was driving alone, whereas the reference transit frequency was a few times a week transit use. The model estimates that for people who do not have mobility restrictions, the log-odds of choosing carpool over driving alone decrease by 0.312 units compared to those who use transit a few times a month. For trip makers having no mobility restrictions, those who use transit daily show a reduced likelihood of choosing carpool mode compared to those who use transit a few times a month. In addition, they express a significantly higher preference for non-motorized modes and transit mode over driving alone. However, travelers without a disability who use transit infrequently or never are less likely to choose transit and non-motorized modes, and they have a stronger inclination towards carpooling compared to those who use transit a few times a week.

Furthermore, for travelers with a disability, daily transit users showcase a substantially increased preference for carpool compared to those reference categories. Additionally, PwDs are more likely to choose transit over driving alone, as supported by ASC, which makes it obvious that disabled people may not drive alone because of increased risks. Interestingly, people having a disability and reporting that they never use transit show a diminished preference towards non-motorized modes with a slight increase in carpool.

## 5. DISCUSSION

This study investigated the factors influencing mode choice behavior among individuals with travel-limiting disabilities. The analysis was based on statistical results curated from over 58,000 trip data. Among the non-disabled population, almost half preferred driving alone, while this percentage dropped to approximately 40% for disabled travelers. In our study, we observed that disabled individuals did not use public transit as expected and favored carpooling, similar to the findings of Schmöcker et al. (2008), where taxi usage increased with age in the presence of a disability and showed a preference for not using public transport.

The MNL models for both groups provided a good fit for all models as described by Louviere et al. (2000). The VOTT for Pw/Ds in Utah was 20.8 cents per minute, indicating a high value placed on time saved during travel, which is similar to a study for work trips in San Fransisco by Koppelman and Bhat (2006) and study in Sydney by Tabasi et al. (2023), although both studies were only for the general population. In contrast, disabled travelers valued time saved lower at six cents per minute, potentially due to their part-time work, free transit fare, and access to shared or no-cost carpooling. This finding supports the study of Lucas et al. (2007) in Honolulu, Hawaii, where elderly people value more transport costs than travel time.

Travelers with disabilities demonstrated a preference for transit and carpool modes over non-motorized options, likely due to mobility difficulties and environmental barriers that discouraged walking. This result may be due to other confounding factors such as employment status, age, and income. With the increase in age, the proportion of walking and biking decreases, which is similar to the finding of Du et al. (2020). The transit time for PwDs is higher as compared to the PwoDs, which makes transit accessibility is significant barrier to choosing transit mode despite being 50% lower fare. It also emphasizes the importance of providing accessible and inclusive transportation options, such as improved public transit services and better connectivity for non-motorized modes, to promote active transportation and enhance the overall well-being of disabled individuals.

Women are more likely to use carpooling in comparison to driving alone across both of the studied groups. The study by Buliung et al. (2009) also delivered a similar result to their study in Canada, specifying that females use more carpools than males. Another study in Spain provided insight that female travel is related to the most sustainable means of transport (Miralles-Guasch et al., 2016). Our disability model shows that women with disabilities take transit more, but those without disability do not. Among working-age adults who do not have disabilities, there is a higher preference for non-motorized modes of transportation over driving. PwDs in the working age showed a negative preference for carpooling as opposed to the Pw/Ds group.

Consistently in both groups, low income is related to increased use of carpooling and non-motorized modes, similar to the results of Kim and Ulfarsson (2004). Moreover, low income is related to increased transit use, and the very-high-income group uses carpool and non-motorized modes less than driving, only among those without disabilities. For very high-income households with disabilities, transit usage is negative compared to driving, suggesting less sensitivity to household income and mode preference in this group. Although disabled people show a preference towards transit use in our modeling, the descriptive statistics results highlighted the reluctance of disabled individuals to use transit, even when it is available at a reduced fee, suggesting a need for policy interventions to promote public transit in Utah. The policy could include accessibility features like easily available transit stations, transit frequency, short and convenient first and last miles, disabled-friendly sidewalks, and proper information systems inside transit vehicles.

Vehicle ownership is not found to be related to a decreased or increased use of non-motorized modes or public transit compared to driving for PwDs, but does so for those without disabilities. Households with disabled individuals owning three or more cars preferred driving alone over carpooling. It might be possible for those who can drive and have other types of disabilities. Households with no car preferred non-motorized modes for non-disabled, whereas counterpart groups were inclined to transit mode. In contrast, the sensitivity for bike availability was lower among disabled individuals, indicating the need for policies to make non-motorized modes more accessible and safer for them.

Furthermore, the study found that Pw/Ds were more likely to use transit over driving alone in downtown/CBD areas, aligning with the perception that expensive parking or congestion in those areas influenced mode choice. However, there can be many other impacting factors, such as walkability and other D variables (Ewing et al., 2019). On the other hand, Non-motorized modes increased in urban areas only for PwDs but not those without disabilities, indicating the importance of the built environment in promoting active and sustainable modes for this vulnerable group.

Regarding age groups, carpooling was more preferred among older individuals with disabilities than working-age adults. This might be attributed to older disabled individuals being more likely to travel with family members or others who can drive for them. When PwDs ride transit daily, they are likelier to use other modes such as carpool, walking and biking. This may indicate that daily transit riders with disabilities tend to have more complex travel patterns involving multiple modes. Likewise, Zhou et al. (2022) reported that elderly individuals heavily rely on public transportation due to physical limitations and restricted car usage. This highlighted the importance of addressing the requirements and preferences of disabled travelers during transit system planning to encourage greater transit utilization. People with no driving licenses show similar and multimodal travel preferences for both studied groups, with most inclined for transit.

Conventional four-step travel demand modeling (TDM) has been used by most MPOs, state departments of transportation, and local planning agencies. The third step of this travel demand gives which mode of travel to use by giving relative shares of available modes. A flaw of the four-step model is the failure to account for the travel behaviors of PwDs. Ewing et al. (2019) reported that more than half of the MPOs in Utah have not included non-motorized modes in their travel demand model. Our study revealed the vital role of the active modes in the mode choice model for both groups and recommended including them in the TDM. Unique trip-making behaviors identified in this study for PwDs may be adjusted in the TDM process from the beginning, i.e., trip generation, to the end, i.e., route assignment, which helps incremental advancement in the state of the practice in traditional TDM. A Few Policy implications can be proposed based on the findings of the travel behavior of PwDs. Improving transport infrastructure, such as public transport networks, increasing transit stations, providing accessible sidewalks, and decreasing first and last trip miles, helps increase transit and non-motorized mode users, which promotes sustainable and equitable transportation.

Our study has limitations, such as extracting trip distances from the BING API to generate mode-specific travel times to the current date. It is essential to consider that travel times for trips with the same origin and destination might have varied due to changes in road network density, transit stops, sidewalks, and other factors. Also, if travel time was disaggregated into access, egress, in-vehicle, and waiting times, the result might have unfolded. In addition, among the three models utilized, the sample size for PwDs was relatively smaller. Due to the uneven distribution of samples among disability attributes, our statistical analysis could be biased. We interpreted the findings of the first two models and the third model with caution.

Future research can explore the potential variations in travel patterns based on different types of disabilities, their severity, and the duration of disability. Since this study lacks detailed disability descriptions, further investigations could explore these aspects. In this study, we focused on residential location and place type as built environmental factors in our model. However, other built environment factors, such as employment density, road network density, and overall area density, might influence the travel behavior of PwDs (Etminani-Ghasrodashti & Ardeshiri, 2016). It is recommended to extend this research by including these additional built environment factors. Moving forward, a study of unobserved heterogeneity and panel effects could be conducted using more sophisticated models like mixed logit models. This model performs with minimal prediction errors in resolving heterogeneity and similarity issues (Chen et al., 2013). This would be a direct extension of our current study.

## 6. CONCLUSIONS

The mode choice models developed in this study using the RP survey are specific to the needs and preferences of PwDs in Utah. A comparison between people with and without disabilities in Utah revealed that disabled individuals have unique travel behaviors. The MNL models showed a good fit for both groups. Our statistical models find a lower value of travel time for people with travel-limiting disabilities than their counterparts, potentially due to multiple factors like part-time work, discounted transit fare, and access to shared or no-cost carpooling. Moreover, our study revealed that despite a 50% fare reduction for PwDs, transit accessibility remains a significant barrier in their choice of Transit mode. Furthermore, an Inclusive mode-choice modeling framework is required to address the unique travel needs of people with disabilities.

The model for PwDs shows that women with disabilities take transit more, but not those without disability. PwDs of working age showed a negative preference for carpooling as opposed to the counterpart group. In both groups, low income is related to increased use of carpooling and non-motorized modes. Vehicle ownership is not found to be related to a decreased or increased use of non-motorized modes or public transit compared to driving for PwDs. Sensitivity to bike availability is found to be lower among disabled individuals. Non-motorized modes increased in urban areas only for PwDs. Daily transit riders with disabilities and non-drivers tend to have more complex travel patterns involving multiple modes.

The study's findings contribute valuable insights into policy implications in inclusive transportation system planning. The distinct travel patterns among individuals with disabilities could be integrated into the entire TDM process, starting from trip generation to route assignment, enhancing the state of practice in traditional TDM. By addressing the unique needs and preferences of disabled travelers, enhancing accessibility and convenience for public transit users, and promoting non-motorized modes, transportation agencies and local governments can work towards a more sustainable and inclusive transportation system. The study's relevance extends beyond Utah and can serve as a representative model for other regions aiming to improve transportation accessibility and inclusivity for PwDs.

**Author Contributions**

**MBK** Conceptualization, Methodology, Validation, Formal analysis, Investigation, Data Curation, Writing - Original Draft, Writing - Review & Editing, Visualization; **ZS** Conceptualization, Methodology, Validation, Investigation, Resources, Writing - Review & Editing, Visualization, Supervision, Project administration, Funding acquisition; **KP** Methodology, Investigation, Writing - Review & Editing, Funding acquisition; **KC** Investigation, Writing - Review & Editing, Project administration, Funding acquisition

**Disclosure Statement**

No conflict of interest

**Acknowledgement**

We would like to thank the multiple anonymous reviewers from TRBAM 2024 for their constructive and insightful feedback, which significantly shaped the development and refinement of this research. Their comments helped strengthen the clarity, rigor, and relevance of our work.

**Funding**

This work was supported by the National Institute on Disability, Independent Living, and Rehabilitation Research [Grant Number 90DPCP0004].

**Data Availability**

A limited dataset would be available upon request to the corresponding author. Some datasets cannot be provided due to data sensitivity, data agreement, and privacy concerns.